%% file: main.tex
\documentclass[conference]{IEEEtran}
\IEEEoverridecommandlockouts
\usepackage{cite}
\usepackage{amsmath,amssymb,amsfonts}
\usepackage{algorithmic}
\usepackage{graphicx}
\usepackage{textcomp}
\usepackage{xcolor}
\def\BibTeX{{\rm B\kern-.05em{\sc i\kern-.025em b}\kern-.08em
    T\kern-.1667em\lower.7ex\hbox{E}\kern-.125emX}}

\usepackage[hyphens]{url}

\begin{document}

\title{Museum Accessibility Through Wi-Fi Indoor Positioning
\thanks{This work was partially supported by Antel.}
}

\author{\IEEEauthorblockN{Antonio Bracco, Federico Grunwald, Agustin Navcevich, Germ\'an Capdehourat and Federico Larroca}
\IEEEauthorblockA{Facultad de Ingenier\'ia, Universidad de la Rep\'ublica\\
Montevideo, Uruguay\\
\{antonio.bracco,federico.grunwald,agustin.navcevich,gcapde,flarroca\}@fing.edu.uy}
}

\maketitle

\begin{abstract}
    \input{abstract.tex}
\end{abstract}


\input{intro.tex}

\input{related.tex}

\input{system.tex}

\input{performance.tex}
\input{conclusions.tex}

\bibliographystyle{IEEEtran}
\bibliography{biblio}

\end{document}

%% file: abstract.tex
Accessibility has long been a primary concern for major museums around the world. This is no exception for the Museo Nacional de Artes Visuales (MNAV, National Museum of Visual Arts) in Uruguay. Having a special interest in achieving accessibility for visually impaired visitors, the MNAV sought to implement a new system to allow these visitors a seamless tour around a new exhibit. We present here the system we developed and the lessons we learned from its deployment and usage. In particular, we used Wi-Fi indoor positioning techniques, so that visually impaired visitors could hear relevant audios through an Android app from their own smartphones based on their location inside the museum. The system was further adapted and used to assist the general public during their visits, allowing access to texts, audios and images according to their position. We furthermore share the complete source code and the dataset used to train the system. 


%% file: intro.tex
\section{Introduction}
\label{sec:intro}

Museums have always looked for new ways to engage their visitors, and in particular address accessibility for people with disabilities such as visually impaired audiences.
Going further than audio guides, diverse solutions have presented themselves as alternatives to making culture more available and engaging. The use of indoor positioning systems have been at the forefront of developing these new systems, allowing a further interaction between user and their position through their smartphone.  These new technologies have the potential to be game-changing in their approach to granting further accessibility and a better experience for these visitors.

In Uruguay, the MNAV (Museo Nacional de Artes Visuales, National Museum of Visual Arts) has been specially committed with accessibility. For instance, in 2015 it deployed a system called ``Museo Amigo'' (Friendly Museum), consisting in a number of totems distributed in front of certain paintings. These totems had a 3D version of the painting for the user to touch as well as a place to put a tablet, provided by the MNAV staff, which would play a specific audio regarding the painting.

Thus, visitors needed to use special tablets that had to be asked from the museum's staff (the system was based on near-field communication) and it also required explanations and assistance on how it was used. These factors were very detrimental to the efforts of making each user independent, making the solution ineffective at engaging visitors. Thus, by the end of 2017, the MNAV was contemplating replacing this system with an alternative for a new exhibit to open on November 2018. 

With this in mind, the authors were contacted to design, deploy and evaluate this alternative, the result of which we discuss in this article.  
In a nutshell, the solution is based around Wi-Fi indoor positioning techniques coupled with an Android app. The main principle behind this idea is that by using the user's position, the app can forgo the use of interactions with the screen (e.g.\ through buttons). Moreover, being based on Wi-Fi, the system may be used by any smartphone or similar device. Finally, the focus on Android was simply because more than 80\% of Uruguayans have this OS on their phones.


More in particular, the visitor can walk around the ground floor of the museum and when nearby a specific artwork, the corresponding descriptive audio will automatically play as an image of the artwork fills the screen of the phone. Although marks on the floor exist to guide the visitor's walk, the person is free to visit the museum as they see fit, which is particularly important for those visually impaired but not completely blind. 

The app was further complemented with a mode designed for the general public. In this mode, those artworks closest to the user are displayed on the phone. By clicking in one of these artworks, the user may read a descriptive text as well as listen to a specific audio about the artwork.

Our main contribution is sharing some valuable lessons we learned from this deployment, in particular regarding the positioning system and its accuracy. 
As we further present in the following section, Wi-Fi indoor positioning systems are implemented through Machine Learning (ML) algorithms that learn to map the RSSI (Received Signal Strength Indicator) observed by the device from several APs (Access Points) to a particular area inside a building. 
This is the so-called \emph{fingerprint}-based scheme, first proposed in the seminal paper by Bahl et al.~\cite{bahl2000radar}. Although several other papers have been published studying this and other indoor positioning systems~\cite{davidson2017survey,zafari2019survey,basri2016survey}, actual deployments are somewhat scarce. The present paper is an effort in this direction, and strives at showing that the technology is perfectly apt for public buildings and massive deployments. 

In what follows, we discuss what level of precision is necessary for the system to be useful, and what it means in terms of number of APs and measurements used to train these ML algorithms (the biggest costs of any new deployment). Some unforeseen problems (and their solution) are also discussed, in particular pertaining to 2.4~GHz-only devices. 
These are all important lessons that we did not find in the literature and we believe will be of interest to the rest of the community. Moreover, all of the software and the complete dataset we used to train and test the system are shared in our repository \url{https://github.com/ffedee7/posifi_mnav}.






%% file: related.tex
\section{Related work}
\label{sec:relwork}



There have been many different approaches when it comes to achieving universal accessibility for museums. Some articles such as~\cite{lisney2013museums,ginley2013museums} explore the problem from the disabled visitors' perspective. 
We would like to highlight The Andy Warhol Museum, which has developed the Out Loud app~\cite{outloud}. This is an inclusive audio guide, which considers different disabilities. In the particular case of blind or low-vision visitors, the app uses an indoor positioning system based on bluetooth low-energy (BLE) beacons to play audios based on visitor's location. Particularly interesting is the fact that it is completely free and the code is available at their repository. 

However, the Out Loud app presents some important disadvantages. Firstly, it is developed for iOS, which restricts the possible audience. As we discussed in Sec.\ \ref{sec:intro}, the possibility for the visitors to bring their own device was a requirement based on the museum's previous experience. Moreover, as the name suggests, these BLE beacons are simply bluetooth emitters with very low power. A phone with bluetooth enabled may then use this signal as an indication that it is near a certain artwork. The most important disadvantage of this positioning system is that several such beacons have to be bought and deployed (roughly one per artwork). Although the price of this hardware is not very high, for museums on a tight budget the best strategy would be to re-use their existing infrastructure: Wi-Fi. 

Although Wi-Fi was not designed for positioning, the APs location is fixed and they periodically broadcast beacon frames from which the AP may be identified (by means of its MAC address). Since the received power (or RSSI) of this beacons depends on the receiver's position, the measurements from several APs may be used to estimate this position~\cite{davidson2017survey,zafari2019survey,basri2016survey}. Although trilateration seems at first as a valid approach to this problem, indoor propagation may be extremely complex, resulting in unpredictable relations between distance and RSSI. 

An alternative is to divide the building into areas (such as rooms, or zones around artworks) and train an ML algorithm to learn to map RSSI readings from available APs to the corresponding area; i.e.\ transform the problem into a classification one. This is the fingerprinting approach to Wi-Fi indoor positioning~\cite{bahl2000radar}. The deployment has thus two stages. First an offline phase where RSSI measurements are obtained for all areas and the ML algorithm is trained. Then an online one where the actual positioning takes place, and the RSSI measurements obtained by the user are fed to the trained algorithms. 
In our final system, we used an ensemble of six machine learning algorithms: support vector machine (SVM), decision trees, random forest, multi-layer perceptron (MLP), AdaBoost and k-nearest neighbors (KNN)~\cite{hastie2009elements}. 
Further discussion and their evaluation are included in Sec.\ \ref{sec:perfeval}. 

Regarding the implementation of the system, the two most prominent open-source alternatives we considered as starting points are Anyplace~\cite{anyplace} and FIND3~\cite{find3}. They implement the most important blocks of an indoor positioning system, which we discuss in the next section. Although they are both valid options, we decided to work with FIND3 since we found its architecture simpler to modify and more flexible (and it has a significantly larger userbase). 

The question that remains is what precision we may obtain from this Wi-Fi positioning system and at what cost. For instance, how many APs are necessary to install or how many measurements should be taken in order to obtain a reasonable precision are important factors to consider in any deployment of such system, and which the available literature does not discuss. Before presenting these issues in Sec.\ \ref{sec:perfeval}, we briefly present our implementation. 






%% file: system.tex
\section{Accessibility solution design }
\label{sec:solucion}

\subsection{On-site design}

Naturally, the more APs are present in the premises, the better the resulting system's precision will be. In particular, the device should ``see'' as many APs as possible in all areas of interest.
 It is important to note that a signal level just above visibility is enough; i.e.\ APs are not required to provide connectivity, which typically required an RSSI above -65/-70\,dBm, but a level above the sensitivity of the devices, typically -90\,dBm, is enough. 

In our particular case, practically speaking the infrastructure was non-existent, so the Wi-Fi network had to be designed and installed from scratch.
Since we wanted to evaluate how many APs were actually necessary, we took the conservative decision to design the system such that in every point of the museum at least 5 APs were visible (more about this topic in Sec.\ \ref{sec:minAPs}).
If the infrastructure is already present, then a site survey is necessary to verify this condition, and check if more APs are necessary (and where).  

\begin{figure}
    \centering
    \includegraphics[width=0.47\textwidth]{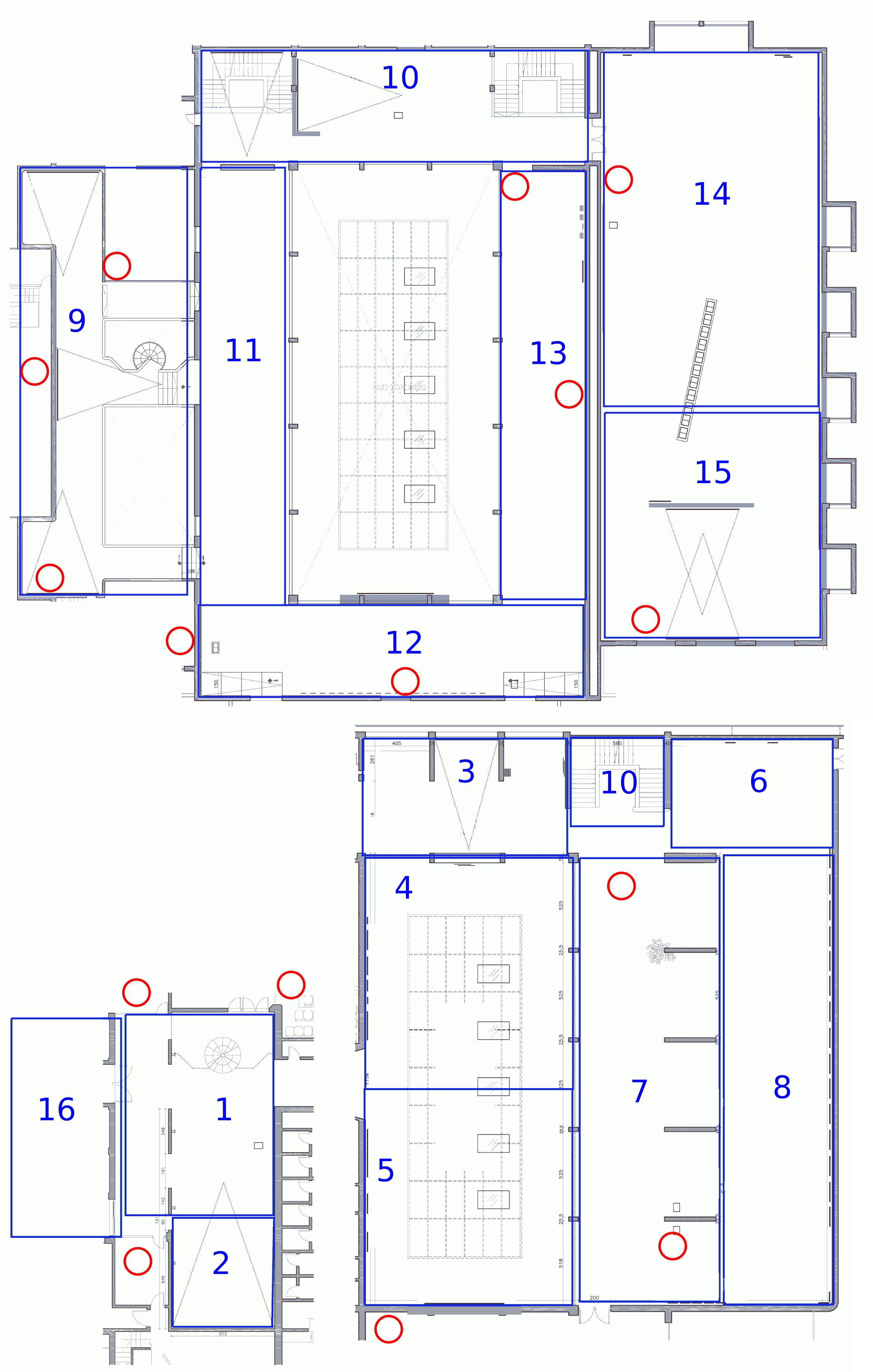}
    \caption{The chosen location areas and their numbering (the lower floor below). Note that zone 16 is outside the building. Red circles correspond to APs. Only areas open the general public are shown.}
    \label{fig:zonas}
\end{figure}

Through a Wi-Fi network design tool, and after some iterations with the museum's authorities, the final disposition includes 15 APs.
Figure \ref{fig:zonas} shows the museum's map (two floors, amounting to about 5000\,square meters in total) along with the AP's final positions (marked as red cirles). In this case, where APs are under our control, both frequency and power were fixed so that the RSSI measured by the devices does not change due to dynamic configurations.  

The next step is to choose what areas will be used for positioning. The final areas chosen in our case and their numbering are also shown in Fig.\ \ref{fig:zonas}. They should be associated with certain artworks, but not too small so as to compromise precision. Moreover, actual physical barriers between areas (e.g.\ walls) naturally help to distinguish among them. For instance, as we discuss in Sec.\ \ref{sec:perf90}, areas numbered 6, 7 and 8 were the most challenging ones, mostly due to the absence of these physical separations. Each of these areas include one of the artworks in the tour for the visually impaired visitors (on the lower floor only)

\subsection{System's architecture}

The proposed system architecture consists of two main parts: a back end, which is responsible for the indoor location, and a front end, that interacts with the user. The back end is on the cloud (in our particular case at Amazon Web Services, AWS) although it may be hosted in any server, and is responsible for the training and execution of the ML algorithms and hosts three databases: one for the training data, another one for the multimedia files and a final one of the user's estimated positions. The latter may be used for analytics on the visitors' habits (hours of visit, popular artworks, etc.). Several important modifications to FIND3 where included in the final version. See the following subsection for details.

The front end part of the system is an Android application. A lot of interdisciplinary work was needed to create the application, regarding aspects ranging from the color palette to the content. Regarding positioning, the app periodically sends to the backend the RSSI of all the APs it senses (once a second by default). The estimated location is returned to the user, and the front end decides whether to update the current area or not. In order to avoid constant changes between areas, we found that a simple rule worked: if the estimated location is three consecutive times the same (and different from the current one), then the front end updates the current area. 

Recall that the application had two modes, depending on the user: general public and visually impaired. In the general public mode it will show one image of each artwork in this new location, and when the user clicks it, it shows an extended screen with more information: text, image and audio. The mode intended for the visually impaired visitors requires almost no interactions. At first the app will play an introductory audio and then it will automatically play audios as the visitor walks among areas. The user can tour the museum as desired and the audios will not reproduce more than one time per session.

An additional and important tool was developed which we highlight here. In order to upload text, images and audios of the artworks to the system a dashboard was created, so the museum staff could do changes by themselves. It is a very simple dashboard where artworks may be created and it is possible to upload media about it. Each artwork is assigned to an area of the positioning system, although an artwork may be mapped to more than one area. After creating an artwork in the dashboard, it will be shown in the Android app when a user is on the area that was assigned to it. Although it may appear almost elementary, it was a key addition to the system in terms of usability.

\subsection{AWS Deployment Details}

\begin{figure}
    \centering
    \includegraphics[width=0.47\textwidth]{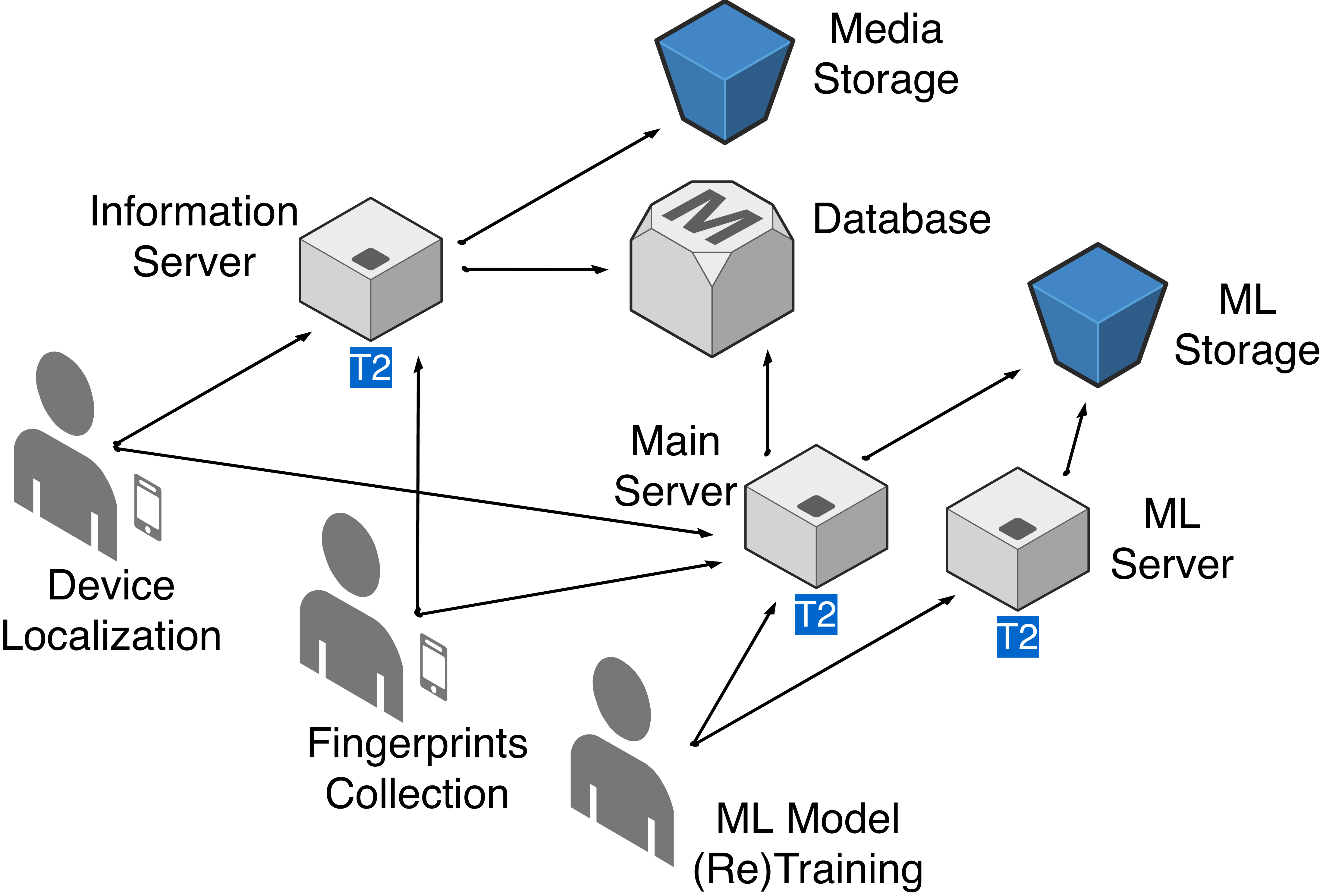}
    \caption{Backend's architecture.}
    \label{fig:arq}
\end{figure}

Figure \ref{fig:arq} shows the detailed implementation of the system in AWS. The three main use cases are depicted in the flowchart. First, the device localization, which corresponds to the standard situation where a user device sends the RSSI measurements of the APs in the area and the system estimates the device location. Then, we have the fingerprints collection process, for which a particular Android app was developped in order to take
the training RSSI measurements and upload them to the system. Finally, we have the training of the ML model, which is only done when new fingerprints are collected. We next briefly describe each of the AWS system architecture components, and highlight the main differences with the FIND3 vanilla system.

\subsubsection*{Main Server}
The main server is hosted on a docker container in a EC2 instance (Amazon Elastic Compute Cloud) of type T2 and medium size (2 vCPUs, 2.3 GHz, Intel Broadwell E5-2686v4, 4 GB RAM). The main server is in charge of receiving all the backend requests, process them and send them to the ML server if necessary. It has direct communication with the database, the ML server and the ML storage. 

Important changes were introduced here with respect to the original FIND3, in particular regarding the communication with the database, which was originally implemented in SQLite~\cite{sqlite}. As we intended the system to scale with several users accessing the database simultaneously, we implemented a separate entity, discussed below.

\subsubsection*{ML Server}
The ML server is responsible for the training and classification of the different algorithms, each time it is requested by the main server. It only receives requests from the main server and also makes use of the ML storage.

\subsubsection*{Information Server}
A different EC2 was used for the information server (type T2 micro, with 1 vCPU, 2.5 GHz, Intel Xeon Family, 1 GB RAM). It provides the artwork information (text, audio and images) for the different museum zones. In the database the different artworks are registered with their corresponding text description and the URLs to the images and audio files.
This component is new with respect to FIND3 and is accessed by the Android app. 

\subsubsection*{Database}
The RDS Database (Amazon Relational Database Service) is an AWS service for relational databases, similar to EC2.
In this case it is used for the implementation of the main database of the system. As we mentioned before, this is a new entity with respect to the original FIND3, and it assumes the database functionalities that are included in the Main Server in the original FIND3. For instance, it stores the labeled fingerprints, prediction results and information about the different zones. 

\subsubsection*{ML Storage}
It is an S3 bucket (Simple Storage Service) which provides an API for fast, flexible and scalable storage for all the ML server data.
The state of the algorithms of the ML model is stored here, as well as a CSV file that contains the data with which the model was trained. This is another new element with respect to FIND3.

\subsubsection*{Media Storage}
Another S3 bucket to store all the media content, such as images and audios for the different artworks. This element was not present either in the original FIND3 implementation.

%% file: performance.tex
\section{Localization performance evaluation}
\label{sec:perfeval}

As discussed before, the positioning problem addressed can be considered as a classification one. With the RSSI-based approach, the goal is to estimate which is the most probable area where the device is located based on a set of RSSI values measured at the device. For this purpose, different standard classification algorithms were used, which were then combined to build a meta-learner using the Youden index~\cite{youden}, as explained next.

First, data is divided into training, validation and test sets. The classic data partition of 70\%, 20\% and 10\% respectively was chosen. After training, for each location $y$ and each algorithm $\omega$ the Youden index $J(\omega,y)$, also known as the informedness statistic and which is equal to
\begin{equation}
J(\omega,y) = sensitivity(\omega,y) + specificity(\omega,y) - 1,
\end{equation}
is computed using the validation data, where
\begin{gather}
Sensitivity = \frac{True~ Positives}{True~Positives + False~Negatives},\\
Specificity = \frac{True~Negatives}{True~Negatives + False~Positives}.
\end{gather}

Given a new RSSI measurement $x$, each algorithm provides a probability $P_{\omega}(y|x)$ for each location $y$. These probabilities are then weighted with the Youden index $J(\omega,y)$ to obtain a total score $Q_y(x)$ that is assigned to each location $y$:
\begin{equation}
    Q_y(x) = \sum_{\omega=1}^{N} J(\omega,y) P_{\omega}(y|x). 
\end{equation}
The location with the highest score is the output of the meta-learner. 

As mentioned before, six machine learning algorithms were used: SVM, decision trees, random forest, MLP, AdaBoost and KNN. All of them are included in the scikit-learn python library~\cite{scikit} used in the FIND3-based system developed. This a subset from all the algorithms included in the original FIND3 implementation. We decided to discard some of them because they did not improve the performance and introduced unnecessary additional computational costs. In any case, this combination strategy proved very beneficial, as none of the algorithms alone obtained better results than the combination. 

To measure the system's performance, we used two metrics computed with the remaining test set. Firstly, the accuracy of the system, defined as the ratio of measurements that were correctly classified. Secondly, the confusion matrix, whose value in the position $i,j$ represents the ratio of data points corresponding to location $i$ that were classified in location $j$. Results shown here correspond to the average after executing the system predictions 10 times. On each execution the data is randomly splitted in the different sets (training, testing and validation).


In the rest of this section, we present the different performance evaluations carried out. As we will see, each result is associated with a lesson learned, which we believe will be useful for future deployments and similar projects.

\subsection{Lesson learned \#1: the accuracy should be above 90\%}
\label{sec:perf90}

During the system setup, around 20,000 RSSI measurements (fingerprints) were collected in the museum, which was divided in the 16 different locations defined on the exhibition map (cf.\ Fig.\ \ref{fig:zonas}).
The number of measurements per location was not uniformly distributed, ranging from 800 to 1500, as more measurements were taken in the most difficult areas (e.g. open spaces with not clear rooms separations). 
Please note that in this case each RSSI measurement has length 30, as we have 15 dual-band APs operating in both frequency bands (2.4~GHz and 5~GHz); i.e.\ we have 2 measurements per AP.

\begin{figure}
    \centering
   \includegraphics[width=0.4\textwidth]{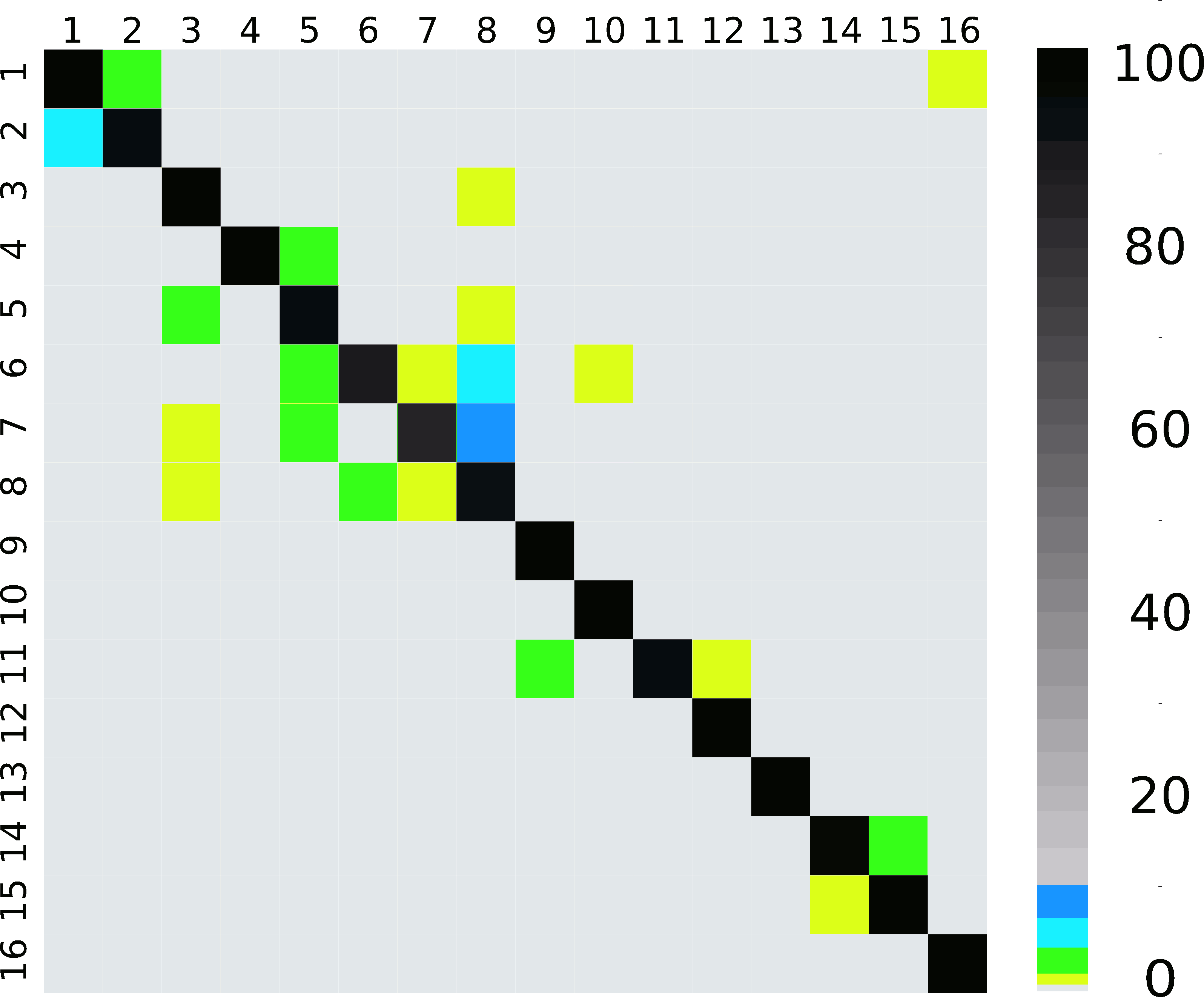}
    \caption{Confusion matrix for the system trained with data from both frequency bands. Values different from zero are highlighted (best viewed in a color display)}
  \label{fig:conf_filtered}
\end{figure}

The resulting overall accuracy was $96.0\%$, whereas Fig.\ \ref{fig:conf_filtered} shows the corresponding confusion matrix. It is worth noting that the values in the diagonal are almost all near 100\%. However, there are a couple of locations which have average values below 90\%, which were the two more problematic museum areas. If we look at locations 6, 7 and 8, we can see some significant confusion in those areas. In practice, those zones were close to each other, so it did not generate major problems concerning the user's experience. However, the lesson learned from this test indicates that a minimum accuracy of 90\% should be ensured to have an appropriate performance in the field.

\subsection{Lesson learned \#2: 2.4GHz-only devices should be treated separately}
\label{sec:only2.4}

One of the first problems we faced when testing the system at the museum was related to the 2.4GHz-only capable smartphones, for which the performance was quite bad.
The problem was that the system did not take into account the user's device. Although certain differences in RSSI measurements among devices are expected, but are very difficult to consider (except by constructing a rich training set), the total absence of 5GHz RSSI measurements is indicative of a 2.4GHz-only device.
This caused that the system, trained with measurements collected with dual band devices, had a very bad performance for devices that are only able to measure in the 2.4GHz band.

To solve this issue, we decided to train and use another classifier for 2.4GHz-only capable devices. The system identifies if the device is dual band or not, just considering the RSSI measurements received, and then it decides which of the classifiers should be used for the location estimation. Each of the RSSI measurements included in the fingerprints are associated with the corresponding MAC address of each AP radio. Then, it is possible to filter the data according to the MAC addresses, in order to select only those radios that correspond to the 2.4GHz band.

\begin{figure}
    \centering
   \includegraphics[width=0.4\textwidth]{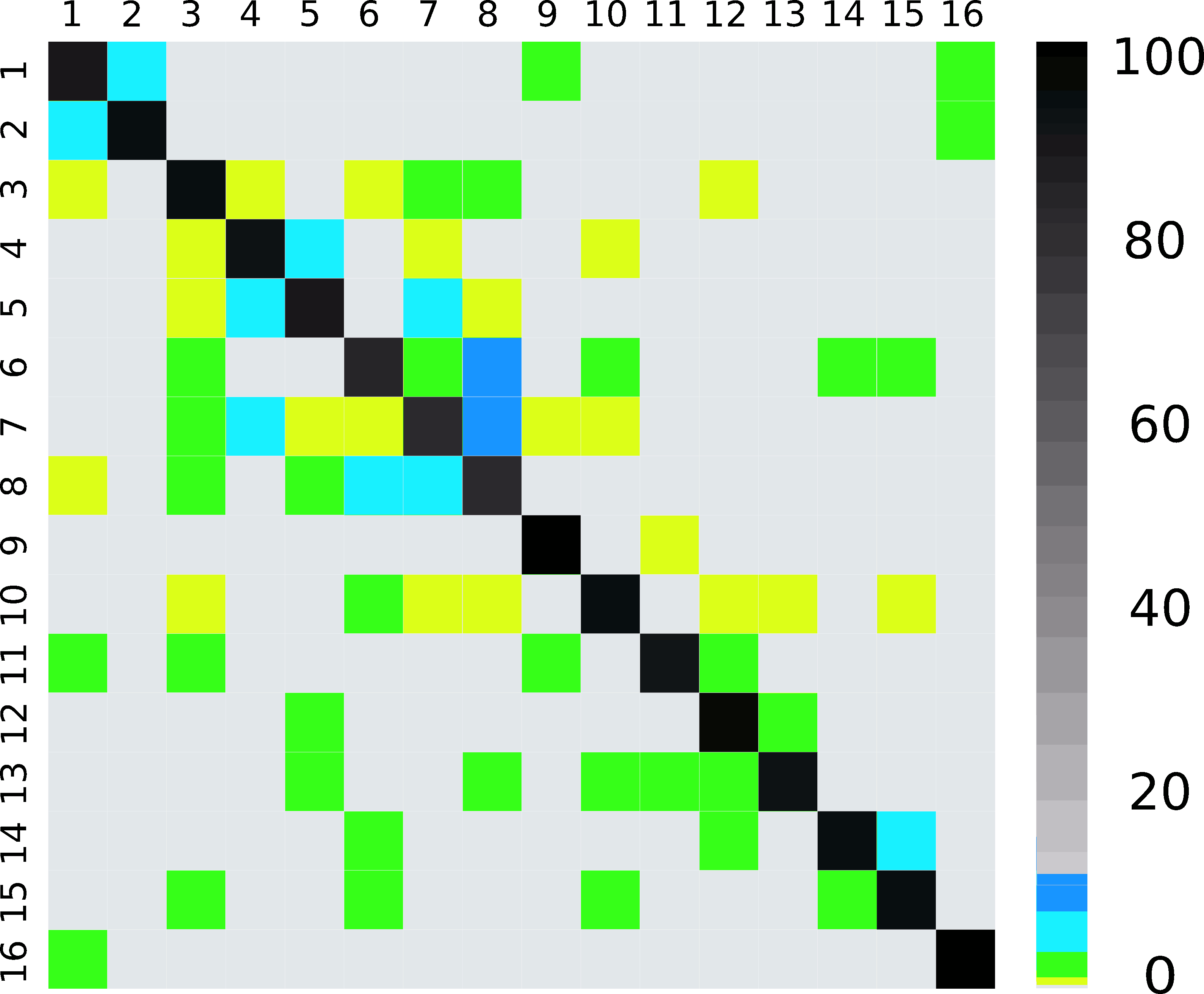}
    \caption{Confusion matrix for the system trained with data from the 2.4~GHz band only. Values different from zero are highlighted (best viewed in a color display)}
  \label{fig:2g}
\end{figure}

Removing the MACs of the radios belonging to the 5GHz band, it was possible to train a new classifier and test the corresponding performance for a 2.4GHz-only capable device. In this case, the average accuracy for the validation set was $90.7\%$. In Figure~\ref{fig:2g} we can see the corresponding confusion matrix, which again shows a good performance for most of the locations, with the vast majority of the classifications lying on the matrix diagonal. We can notice again some problems at locations 6, 7 and 8, where most of the confusions occur.

Concerning the system deployment, the proposed solution was implemented by restructuring the framework code. As previously mentioned, for each location estimation the system decides if the device is 2.4GHz only based on the RSSI measurements. If it has measurements for both frequency bands one classifier is used, and if it only has measurements for the 2.4GHz band the other one is applied. The additional computational costs and delay for the estimation is negligible. This way we obtained a good solution to the problem, achieving a similar behavior in terms of performance for dual band and 2.4GHz-only capable devices.

\subsection{Lesson learned \#3: each location should be covered by at least 3 APs}
\label{sec:minAPs}

Next, we will analyze how the number of APs affects the system's performance. This study is of great importance in terms of cost, bearing in mind that the purchase and installation of the APs (if necessary) is perhaps the most expensive aspect for the system deployment. On the other hand, there is a clear relationship between the number of APs and the accuracy of the system. So, we discuss now how many APs are actually needed to achieve a reasonable system performance.

For this purpose, we follow the same system evaluation procedure described above, but varying the number of APs. 
To discard APs we proceeded with the criterion of eliminating the most redundants in signal coverage first. This way, it is possible to emulate the scenario where the installation was initially planned with a smaller number of APs. In practice, the data for each AP was removed by simply taking into account the corresponding MAC address of its radios and filtering the data from the RSSI fingerprints.

\begin{figure}
    \centering
  \includegraphics[width=0.35\textwidth]{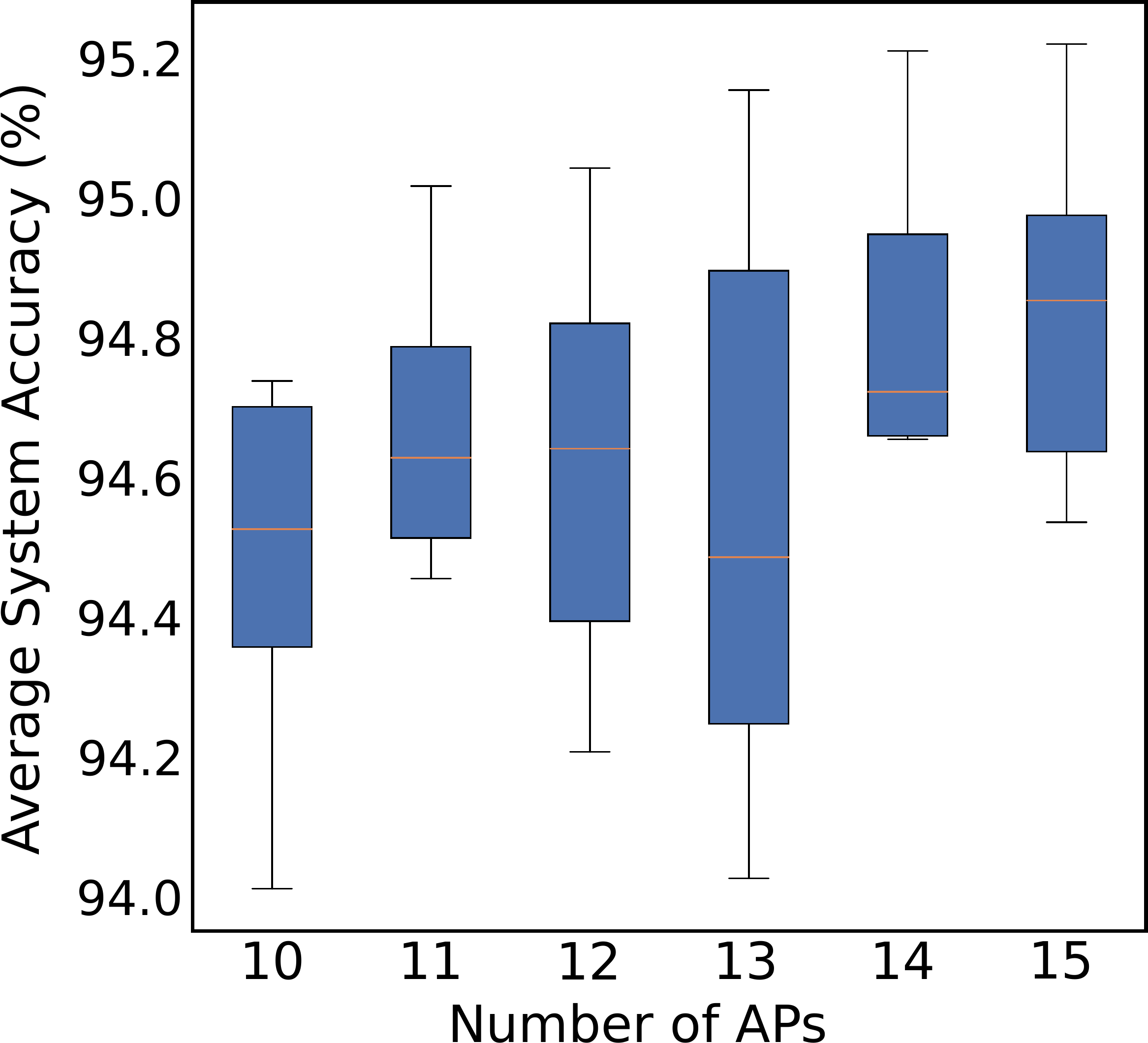}
  \caption{Average system accuracy for the different number of APs. The boxplots correspond to the ten performance evaluations.}
  \label{fig:cant_aps}
\end{figure}

Figure~\ref{fig:cant_aps} shows the average system accuracy for the different number of APs. For each case, we have ten different performance evaluation results, so the quantiles 25\% and 75\%, and the minimum and maximum values are also indicated. For the worst case analyzed we have an average system accuracy of $94.5\%$, which corresponds to the case of a deployment with 10 APs. Most importantly, analyzing the fingerprints, we have verified that it corresponds to having coverage from at least three different APs at each location. This amounts to six measurements (since each AP is dual-band) per location, and it is a very important rule-of-thumb for future deployments. 


\subsection{Lesson learned \#4: the minimum number of fingerprints collected for each location should be 500}
\label{sec:minFing}

Considering the deployment costs, the process of collecting all the fingerprints at each location is another important issue in terms of the working hours. For example, the application we designed and used to collect fingerprints gathers a new measurement every 2 seconds. This means roughly 12 hours to collect the 20,000 fingerprints, without counting some iterations that were necessary to reinforce certain zones. As in the previous case of the number of APs, there is a clear relationship between the number of fingerprints for each location and the corresponding system performance~\cite{zafari2019survey}. Thus, in this section we analyze which is the minimum number of fingerprints needed to reach a reasonable system performance.

For this analysis, the average accuracy of the system was calculated for different number of fingerprints. Starting from the complete set of 20,000 fingerprints, random subsamples were applied in order to obtain fingerprints sets of different sizes, ranging from $30\%$ to $100\%$ of all the measurements. In order to apply the subsampling, the following rules were taken into account:
\begin{itemize}
    \item The proportion of fingerprints for each location is the same, and it is equal to the corresponding percentage sampled from the total number of fingerprints. This way, all the locations keep the same ratio of fingerprints that they have in the complete fingerprints dataset.
    \item The selected fingerprints are chosen randomly for each performance evaluation test. 
    \item Due to the random nature affecting the choice of the subsets of fingerprints, the accuracy computation is calculated over 10 choices.
\end{itemize}

\begin{figure}
    \centering
  \includegraphics[width=0.35\textwidth]{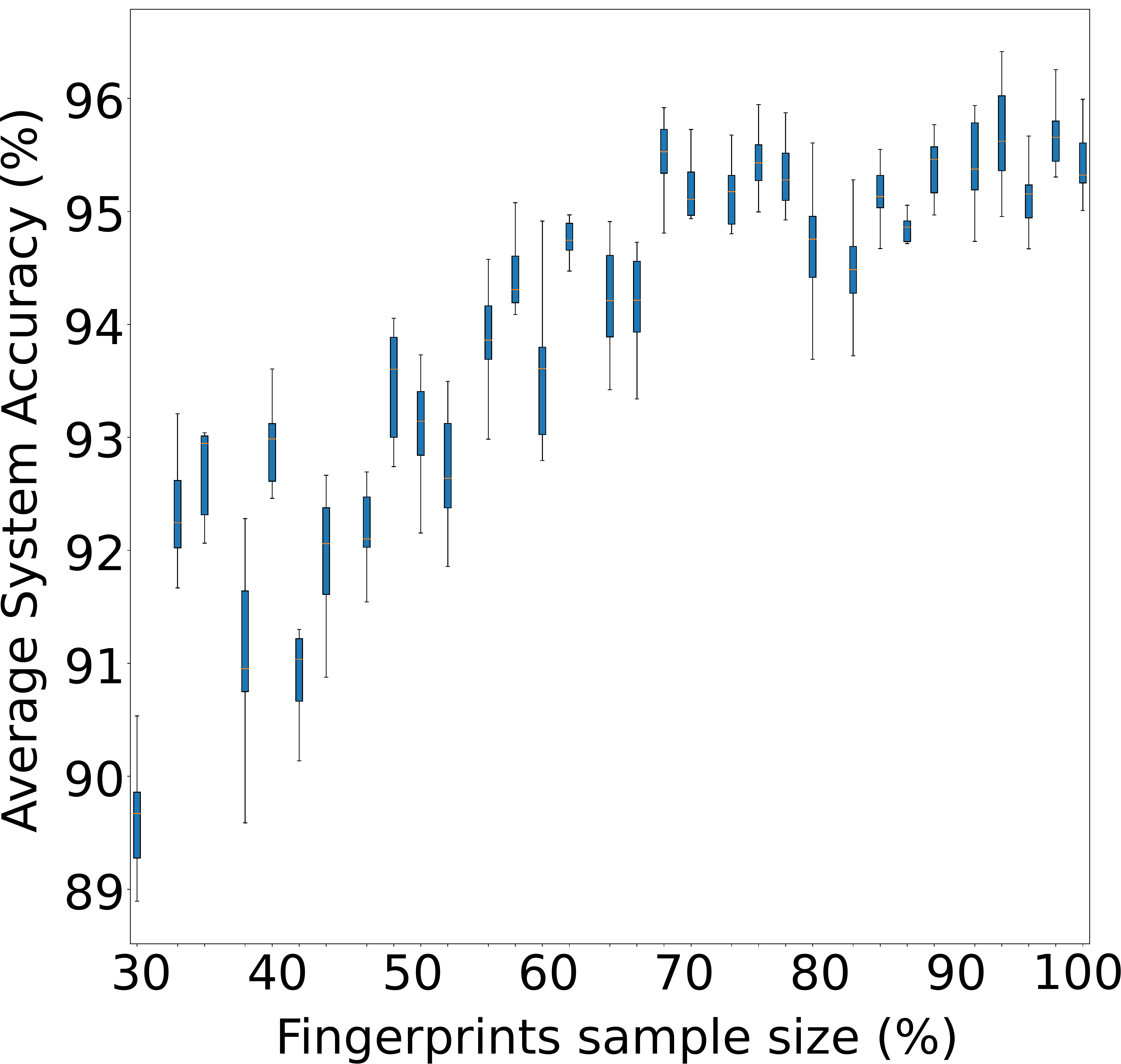}
  \caption{Average system accuracy for the different number of fingerprints. The boxplots correspond to the ten performance evaluations.}
  \label{fig:cant_fingers}
\end{figure}

In Figure~\ref{fig:cant_fingers} we can see the system accuracy as the number of fingerprints varies, which shows a clear tendency to decrease as the number of fingerprints considered is lower. In the same way as in the previous analysis for the number of APs, we look for an appropriate minimum number of fingerprints to ensure a good system performance. Considering that the average accuracy should be above $90\%$, looking at the curve it indicates that at least 40\% of the fingerprints are required (roughly 7,000 fingerprints), for the 16 locations defined at the museum. Recalling that the number of fingerprints measured by location varied from 800 to 1,500, we can conclude that a minimum of 500 fingerprints per location should be considered as a general rule for similar deployments.

%% file: conclusions.tex
\section{Conclusions and future work}
\label{sec:conclusions}

A solution was designed and implemented to enable visually impaired visitors to have a better user experience on their tour through an art museum, in this case the MNAV in Uruguay. The system and application developed, based on Wi-Fi indoor positioning techniques, proved to be a successful solution~\cite{observador,ministerio}, achieving good performance to provide users an interactive experience on their visit to the museum exhibit.
All  the  software developed and  the complete dataset, are publicly available in our repository \url{https://github.com/ffedee7/posifi_mnav}.

Although Wi-Fi based positioning has accumulated several years of research and an important literature is available (see for instance the very recent survey~\cite{zafari2019survey}), its usage is still not very extended, with few institutions using solutions based on them around the world. Our work's main contribution is precisely in this direction: showing that these technologies already have the maturity necessary for massive deployments. We evaluated a system based on the most popular open-source indoor positioning framework~\cite{find3} and share with the community important lessons we have learned in the process that will prove useful for future deployments. 


In particular, the localization performance evaluation and the user experience we have surveyed has shown that the average accuracy should be above 90\% to provide a successful user experience. In addition, the trade-off between the system performance, the number of APs and the number of training measurements were analyzed, being both factors strong influences in the deployment costs. The results show that ensuring the coverage of 3 APs everywhere in the building, and taking approximately 500 training measurements per location, should be enough to have a good performance in most cases.

Analyzing possible extensions to the system, it is clear that the use of alternatives such as BLE could be useful, in particular to improve the spatial granularity of the localization. The indoor positioning based on Wi-Fi measurements has limited capabilities, only enabling to identify the room or a broad area where the device is located. In this sense, BLE beacons present advantages when a high density of beacons is deployed. Combining BLE with the Wi-Fi based solution will improve the performance over short distances. This would allow for example to display content when the user is exactly in front of an artwork. In fact, our system may integrate BLE measurements if present, and we are currently starting to experiment with this technology as a complement to the present deployment. 

Another possibility would be to use some of the several sensors that are commonly integrated nowadays in most end user devices. For example, the accelerometer and the gyroscope could also be useful to improve the localization. They can also be helpful to identify for example where the user is looking at. For this purpose the camera could also be useful, also allowing the artworks' recognition with a suitable previously trained algorithm. A problem with most of these sensors is that they are typically not very accurate and there is a huge variability among different devices, so the integration could be quite challenging.